\documentclass{iaus}
\usepackage{graphicx}

\title[Gaia data processing] 
{Gaia: Organisation and challenges for the data processing}

\author[F. Mignard]   
{F. Mignard$^1$, C. Bailer Jones$^2$, U. Bastian$^3$, R. Drimmel$^4$,
L. Eyer$^5$, D. Katz$^6$, F. van Leeuwen$^7$, X. Luri$^8$, W.
O'Mullane$^9$, X. Passot$^{10}$, D. Pourbaix$^{11}$ \and T.
Prusti$^{12}$}
\affiliation{
$^1$ Observatoire de la C{\^o}te d'Azur, Nice, France \\ email: {\tt francois.mignard@oca.eu} \\[\affilskip]
$^2$ Max Planck Institut f{\"u}r Astronomie, Heidelberg, Germany\\[\affilskip]
$^3$ Astronomisches Recheninstitut (ARI), Heidelberg, Germany\\[\affilskip]
$^4$ Osservatorio Astronomico di Torino, Torino, Italy\\[\affilskip]
$^5$ Observatoire de l'Universit{\'e} de Gen{\`e}ve, Sauverny, Switzerland\\[\affilskip]
$^6$ Observatoire de Paris-Meudon, Meudon, France\\[\affilskip]
$^7$ Institute of Astronomy, Cambridge, England\\[\affilskip]
$^8$ University of Barcelona, Barcelona, Spain\\[\affilskip]
$^9$ European Space Astronomy Centre, Madrid, Spain\\[\affilskip]
$^{10}$ Centre National d'Etudes Spatiales (CNES), Toulouse, France\\[\affilskip]
$^{11}$ Universit{\'e} Libre de Bruxelles, Brussels, Belgium\\[\affilskip]
$^{12}$ European Space Agency, ESTEC, Noordwijk, The Netherlands
}

\pubyear{2008}
\volume{IAU Symposium 248}  
\pagerange{1-7}
 \jname{Title of your IAU Symposium}
\editors{A.C. Editor, B.D. Editor \& C.E. Editor, eds.}
\begin{document}

\maketitle

\begin{abstract}
Gaia is an ambitious space astrometry mission of ESA with a
main objective to map the sky in astrometry and photometry down to a magnitude 20 by the end of the next decade.
While the mission is built and operated by ESA and an industrial consortium, the data processing is entrusted to
a consortium formed by the scientific community, which was formed in 2006 and formally selected by ESA one year later.
The satellite will downlink around 100 TB
of raw telemetry data over a mission duration of 5 years from which a very complex
 iterative processing will lead  to the final science output: astrometry with a
final accuracy  of a few tens of microarcseconds, epoch photometry in wide and narrow bands,
 radial velocity and spectra for the stars brighter than 17 mag.
We  discuss the general principles and main difficulties of this
very large data processing and present the  organisation of the
European Consortium responsible for its design and implementation.

\end{abstract}
\firstsection 
\section{Introduction}
The ESA mission Gaia is a powerful astronomical space project
dedicated to high precision astrometry, photometry and
spectroscopy. The details of the science objectives together with
the principles of the measurements are given in a companion paper
(\cite{gst07}) in this proceeding and are not repeated.  Gaia will
survey the whole sky and detect any sufficiently point-like source
brighter than the 20th magnitude. The observations are carried out
with a scanning instrument mapping the instantaneous field of view
(composed of two disconnected areas on the sky) onto a single
mosaic of CCD very precisely mounted on the focal plane. Therefore
each detectable source is observed during its  motion on the focal
plane brought about by the spin of the satellite. On the average a
source will transit about 80 times during the five years of the
observing mission, leading to about 1000 individual observations
per object. They will be more or less regularly distributed over
about $\sim 40$ epochs with about one cluster of $\sim 25$
individual observations every 6 weeks. These observations consist
primarily of a 1D accurate determination of the image location
at the transit time on a frame rigidly attached to the payload
together with an estimate of the source brightness. The number of
sources potentially observable is of the order of one billion,
primarily stars. In addition there will be something like $\sim
5\times 10^5$ QSOs and $\sim 3\times 10^5$ solar system objects.

\section{Challenges with the data processing}
 Scientifically
valuable information gathered by Gaia during its lifetime will be
encased in the nearly continuous data stream resulting from the
collection of photons in the approximately 100 on-board CCDs in the
astrometric, photometric and spectroscopic fields of Gaia. However
in its original telemetry format the data is totally unintelligible
to scientists, not only because it is squeezed into packets by the
numerical coding but, more significantly, because of the way  Gaia
scans the sky. Indeed with a 1D measurement at each source transits,
Gaia picks up tiny fragments of the astrometric parameters  of the
one billion  observable sources in the sky. To translate this
information into positional and physical parameters expected by
scientists a large and complex data analysis must be implemented.
How huge and how complex it is, is briefly addressed in this
section.

\subsection{Volume of data} Data collected by the Gaia detectors
will be transmitted  by the spacecraft from its position around L2
at a few Mbps for at least five years. Over its lifetime the
satellite will deliver to the ground station an uncompressed raw
data set of approximately 100\,TB, primarily composed of CCD
counts collected during the transit of celestial sources on the
detector. This will make up the raw data. However the data
processing will generate numerous intermediate data to produce the
scientifically meaningful data products by 2020. Given these
intermediate steps, the need to store the provisional solutions
and the different backups, it is estimated that the actual size of
the Gaia related data will be in the range of the PB ($10^{15}$
Bytes), still a fairly large volume in today standard, not yet
matched by any other space mission, although ground based
experiments in particle physics will produce data volume one or
two orders of magnitude larger. It is enough to say that to store
the text of all the books published every year a storage space of
about 1TB is sufficient which could be readily available on anyone's
desk. The orderly access to this data is another question. For the
astronomers accustomed to working with the CDS on-line data bases,
all the catalogues stored in the VizieR service make less than 1
TB in disk space, while the Aladin sky atlas with the images is
just above 5 TB. Despite this apparently limited volume, the
administration and maintenance of these services require several
permanent staff. Therefore volume alone is not the major issue,
while efficient access to the relevant piece of data is a serious
one.

The core of the final data products distributed to the scientific
community will be much smaller. Scaling from Hipparcos with $10^5$
sources with all the results on a few CDs, this should amount for Gaia
to something around 20 TB of data. This depends very much on whether
raw or nearly raw data are included in the delivery. If one
considers just the final data of astrometry, photometry and
spectroscopy (not the spectra but their exploitation in term of
radial velocity and astrophysical diagnostic), all the Gaia data
expected for a standard user can be estimated by considering a data
base with $10^9$ entries, each of these entries comprising less than
$100$ fields. The astrometric data will need many significant digits
(or bytes), but most of the other fields will require a limited
number of digits or bits. A volume of 1-3 TB should then be
sufficient for the main Gaia product, not including  data produced
during the intermediate steps of the data processing. This data will
certainly be made available with calibration data to allow specific
reprocessing with improved models by users.

\subsection{Computational complexity} There is not an all-purpose
definition of what a complex computation is. It could be complex
because of the number of elementary operations it requires (time
complexity),  because of the data management (interface
complexity), or because of the size of the core memory  needed to
perform some parts (space complexity). Gaia can be considered
complex on at least two counts:

 \vspace{5pt}
\begin{itemize}
    \item  It implies a very large resource of CPU, or more or less
    equivalently, the number of basic arithmetical operations is
    very large by today's standards, and will remain large in the
    coming ten years. \\
    \item The data is widely interconnected with access sometimes
    to subset arranged in chronological order (observations of a
    certain type in a certain time interval) or by sources (all
    the observations of a subset of sources).  This extends to the
    access to intermediate data or to calibration files.
\end{itemize}
\vspace{5pt}

 What is at stake can be easily grasped if we consider that there
will be about $10^9$ individual sources to be processed and just
spending one second on each to get the calibrations, spacecraft
attitude, astrometric parameters, multi-color photometry and
spectra would amount to 30 years of computation. Obviously the
overall system must be very efficient to have a chance to produce
the scientific results no less than three years after the end of
the operations.

As to the actual computation size it is not easy to assess with
reliability within at least one order of magnitude as shown
already in 2004 by M. Perryman (\cite{perryflops}). However we
have now a better view  with scalable experiments on simulated
data. The core iterative processing (referred to as AGIS) needs
have been evaluated by Lammers (\cite{agis}) to $4\times 10^{19}$
FLOPs for the last cycle, when all the observation material is
available. Allowing for the intermediate cycles this yields  $2
\times 10^{20}$ FLOPs for the astrometric processing once the
image parameters task and source matching are completed. This
latter processing has also been evaluated from realistic simulated
telemetry and should not be larger than $4 \times 10^{18}$ FLOPs
for the full telemetry volume. Regarding the other processing
(complex sources, photometry, spectroscopy and before all the
image updating) the uncertainty remains significant and depends
also on how much additional modeling will be required to allow for
the radiation effect on the CCDs. A conservative estimate
indicates that the most demanding task will be the image updating
(improvement of  the image parameters and initial centroiding once
a better attitude and calibration are known). It is estimated to
be of the order of  $10^{21}$ FLOPs. But with numerous small
computations repeated on all the sources instead of a global
adjustment requiring big core memory access, this demanding
subsystem could be either distributed or implemented with  a
parallel architecture.

\begin{figure}[h!tb]
\begin{center}
 \includegraphics[width=11cm]{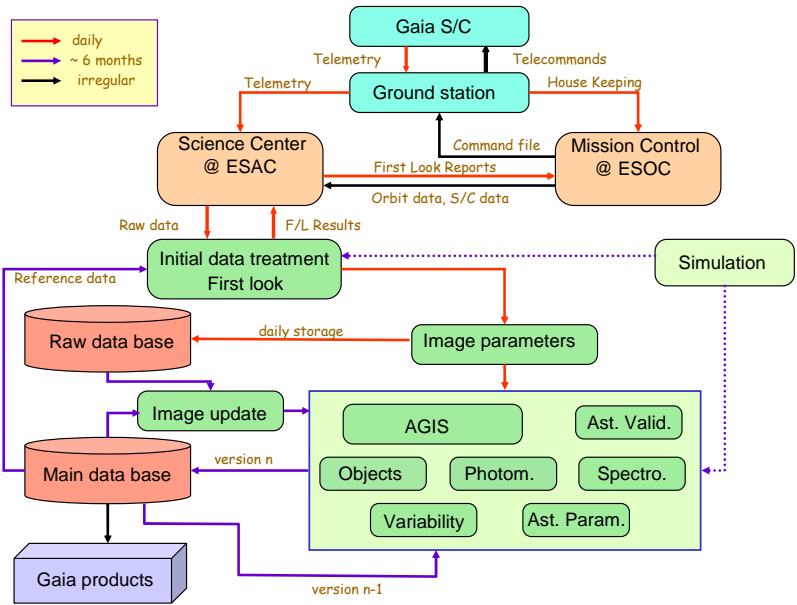}
 \caption{Main structure of the data flow in the  processing of the Gaia raw data.}
   \label{fig:dataflow}
\end{center}
\end{figure}

This figure of $10^{21}$ FLOPs happens to be very close to the
largest computations undertaken so far, like the search of the
largest prime numbers or the location of the non trivial zeros of
the Riemann $\zeta$ function.  The recent distributed computation
for the protein structure (\cite{fah})  is also in the same area in
term of computing size. However these computations are much
different form Gaia's, since they are distributed through internet
on several thousands processing units of small power (and cheap) and
they don't have significant data handling. They are pure
computations where each unit is given the task of making a tiny
contribution. Most of the Gaia computations cannot be distributed in
this way and each computing center will have to access  a large
amount of data with a significant I/O overhead.

By the time Gaia flies the computing power available to the Gaia
Data Processing Centers is expected to be close to 10-20 TFLOP/s
on large clusters. In addition one can also count on the
part-time availability of the Barcelona Super Computer (Mare
Nostrum) which has a today power of 20 TFLOP/s. A straight
evaluation for Gaia tells that about 2 full years of CPU at 20
TFLOP/s will be sufficient to process the data. This is still
daunting but tractable, provided the estimate is not wrong by two
orders of magnitude.

\section{Overview of the Data Processing}

 The Gaia data processing is based on an iterative
self-calibrating procedure where the astrometric core solution is
the single largest treatment and the cornerstone of the whole
processing. These iterations reflect the fact that, as a survey
satellite reaching potential accuracy and completeness levels
never obtained before, it has to be self-calibrating in nearly all
aspects of its observations. The most obvious in this respect are
the astrometric data: the same data that will ultimately form the
astrometric catalogue are also used to reconstruct the scan-phase
of the satellite (along-scan attitude), which is the reference
frame for the astrometric measurements. Similar considerations
apply to the spectroscopic and photometric data. Additional
complications arise from the fact that these three main iterative
solutions are in various ways interlinked and rely on
common calibration data.

The overall processing starts with an initial data treatment
aiming to obtain a first estimate of the star positions and
magnitudes together with an initial attitude, primarily based on
the on-board crude determination improved on the ground with more
accurate (and more computer greedy) algorithms.  The next step  is
the iterative astrometric core solution which provides the calibration
data and attitude solution needed for all the other treatments, in
addition to the astrometric solution of about 100 million primary
sources and the rigid reference system.  Once these two steps are
performed and their products stored in the central data base, the
more specialised tasks are launched with the photometric
processing and variability detection, the global analysis of the
spectroscopic data and the data treatment of all the difficult
sources, like planets and multiple stars, that do not fit in the
general astrometric or photometric solution of single stars. Two
more processing steps close the chain, dealing with the analysis of all
the types  of variable stars and the retrieval of the stellar
astrophysical parameters like their luminosity, temperature or
chemical composition. Each step in the processing using earlier
data has it own logic regarding the products it will deliver, but
must also be seen as part of the validation of the upstream
processes. The overall data flow from the telemetry to the final
products is sketched in Fig.~\ref{fig:dataflow}. The daily flow
follows naturally from the data downlink taking place once per day
during the visibility of L2 above the ESA main station in Spain.

The data exchange between the different processing centers (DPCs)
takes place via a Main Database (MDB) hosted at a single place.
Intermediate processed data from  the processing centers flow into
the MDB; the processing centers can then extract the data from
this Database for further processing. Thus the Main Database acts
as the hub the Gaia data processing system and must have an
efficient and reliable connection with all the DPCs. During
operations, the plan is to version this database at regular
intervals, typically every six months, corresponding roughly to
the period for the satellite to scan the whole celestial sphere.
In this way each version of the Main Database is derived from the
data in the previous version, supplemented with the processing of
the newest observations.

\begin{table}[bth]
\setlength{\tabcolsep}{3mm}
  \begin{center}
  \caption{Coordination Units of the DPAC and their current leader.} \vspace{5pt}
  \label{cus}
 {
  \begin{tabular}{c l l l l}\hline
{\bf CU} & {\bf Name} & {\bf Leader} & {\bf Affiliation} & {\bf Location} \\[2pt]
\hline
 CU1 & System Architecture       & W. O'Mullane    & ESAC & Madrid \\[4pt]
 CU2 & Simulation                & X. Luri         & UB   & Barcelona \\[4pt]
 CU3 & Core processing           & U. Bastian      & ARI  & Heidelberg \\[4pt]
 CU4 & Object processing         & D. Pourbaix     & ULB  &  Brussels \\[4pt]
 CU5 & Photometric processing    & F. van Leeuwen  & IOAC & Cambridge   \\[4pt]
 CU6 & Spectroscopic processing  & D. Katz         & OBSPM & Paris  \\[4pt]
 CU7 & Variability processing    & L. Eyer        & Obs. Geneva & Geneva \\[4pt]
 CU8 & Astrophysical parameters  & C. Bailer-Jones & MPIA & Heidelberg \\
  \hline
  \end{tabular}
  }
 \end{center}
\end{table}

\section{Organisation of the Scientific Community}
To cope with the processing challenge, the scientific community, together
with ESA, has set up a data processing ground segment comprising a
single processing system (no overall duplication) which will deliver  the intermediate
and final mission science products. Since mission selection, the
underlying principles of the data processing have been developed
by the Gaia scientific community and individual pieces were
successfully tested on small or intermediate size simulations.
During this phase one has attempted to identify the critical
elements of this processing (size, iterative procedures,
instrument calibration, data exchange, human and financial
resources, computing power) and assess the risks inherent to an
endeavour of that size, unprecedented in astronomy.

 Based on these
preparatory activities the community has joined forces into a
dedicated consortium: the Data Processing and Analysis Consortium
(the DPAC). In short, the DPAC is a European collaboration
including the ESA Gaia Science Operations Centre (SOC) and a
broad, international science community of over 320 individuals,
distributed on more than 15 countries, and including six large
Data Processing Centres (DPCs, Table~\ref{dpcs}). The Consortium has carefully
estimated the effort required and has united in a single
organisation the material, financial and human resources, plus
appropriate expertise, needed to conduct this processing to its
completion in around 2020.  In 2006 the DPAC has proposed to ESA a
complete data processing system capable of handling the full size
and complexity of the real data within the tight schedule of the
mission. The details of this system, its expected performances,
funding, organisation and management  are described in a document
submitted to ESA as a Response to its Announcement of Opportunity
for the Gaia data processing (\cite{response}).

\begin{figure}[htb]
\begin{center}
 \includegraphics[width=12cm]{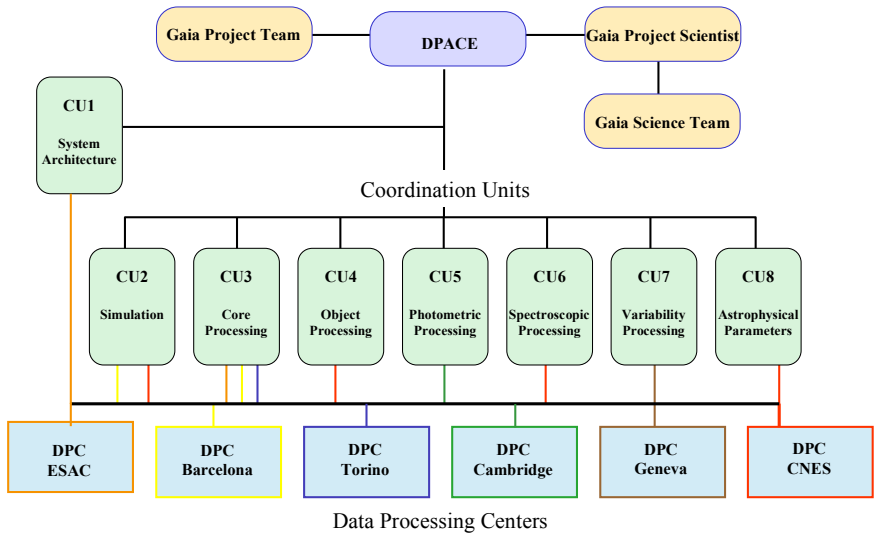}
 \caption{Top level structure of the DPAC with the Coordination Units (CUs) and the data processing centers (DPCs). Colored connectors indicate the link between the CUs and the DPCs.  }
   \label{fig:dpac}
\end{center}
\end{figure}

The Consortium is structured around a set of eight Coordination
Units (CUs) each in charge of a specific aspect of the data
processing. The CUs are the building blocks of the Gaia Data
Processing and Analysis Consortium and they are listed together
with their leader's name in Table~\ref{cus}. An additional CU (the
CU9) in charge of developing the tools for the Catalogue access by
the scientific community is planned in the near future but not yet
activated. The CUs have clearly-defined responsibilities and
interfaces, and their boundaries match naturally with the main
relationships between tasks and the associated data flow.
Responsibilities of the coordination units include: (a) defining
data processing tasks and assigning responsibilities; (b)
establishing development priorities; (c) optimizing, testing and
implementing algorithms; (d) verifying the quality of the science
products. Each coordination unit is headed by a scientific manager
(the CU leader) assisted by one or two deputies and, where
appropriate, a technical manager. The management team of each CU
is responsible for acquiring and managing the resources needed for
their activities. While the CUs are primarily structured for
software development, all of them are closely associated with at
least one DPC where the algorithms will be executed for the
data processing in the operational phases.

The Consortium is coordinated by the `Data Processing and Analysis
Consortium Executive' (DPACE) committee.  This top-level management
structure deals with matters that are not specific to the internal
management of a CU, defining standards and policies needed to ensure
an efficient interaction between all the CUs. Consistent with the
Science Management Plan, the DPACE and its chair will serve as the
interface between the DPAC and the Project Scientist and the Gaia
Science Team. They are ultimately responsible for the data
processing carried out by the DPAC. This executive committee is
composed  at the moment of the DPAC chair and deputy, the leaders of
each CU, a representative of the CNES Data Processing Center. The
Gaia Project Scientist (an ESA position) has a standing invitation
to the DPACE where he has the status of observer.

\begin{table}[bth]
\setlength{\tabcolsep}{4mm}
  \begin{center}
  \caption{The Data Processing Centres associated to the DPAC, their current manager and involvement.} \vspace{5pt}
  \label{dpcs}
 {
  \begin{tabular}{c l l l }\hline
{\bf DPC} & {\bf Location} & {\bf Manager} & {\bf Linked to:} \\[2pt]
\hline
 DPC-B & Barcelona      & S. Girona        & CU2, CU3  \\[4pt]
 DPC-C & CNES           & X. Passot        & CU2, CU4, CU6, CU8 \\[4pt]
 DPC-E & ESAC           & J. Hoar          & CU1, CU3 \\[4pt]
 DPC-G & Geneva         & M. Beck          & CU7 \\[4pt]
 DPC-I & Cambridge      & F. van Leeuwen   & CU5  \\[4pt]
 DPC-T & Torino         & A. Volpicelli    & CU3  \\
  \hline
  \end{tabular}
  }
 \end{center}
\end{table}

The DPAC has responded to the Announcement of Opportunity released
by ESA on 9 November 2006. The Response document contains, in an
hefty volume of more than 650 pages, the overall description of the
Gaia data processing to reach the scientific objectives, the
organisation of the consortium and the funding commitments of the
national agencies supporting the DPAC. This response has been
reviewed by various ESA advisory committees and after one
iteration the DPAC proposal and its selection have been formally
endorsed by the ESA Science Program Committee in its meeting of
May 2007. In  November 2007 the same high level committee has also
approved the funding agreement between the national agencies and
ESA.

\section*{Acknowledgement}
This short review of the DPAC organisation and activities relies
heavily on the contribution of many  members of the DPAC who  are
collectively gratefully acknowledged for their dedication to the
project.

\end{document}